\documentclass[
 aps,
 amsmath,amssymb,
]{revtex4-1}

\usepackage{graphicx}
\usepackage{dcolumn}
\usepackage{bm}

\begin{document}


\title{Interacting superradiance samples: modified intensities and timescales, and frequency shifts}

\author{Martin Houde} \email{mhoude2@uwo.ca}
\affiliation{Department of Physics and Astronomy, The University of Western Ontario, London, ON, N6A 3K7, Canada}

\author{Fereshteh Rajabi} \email{f3rajabi@uwaterloo.ca}
\affiliation{Institute for Quantum Computing, University of Waterloo, Waterloo, ON, N2L 3G1, Canada}

\date{\today}
\begin{abstract}
We consider the interaction between distinct superradiance (SR) systems and use the dressed state formalism to solve the case of two interacting two-atom SR samples at resonance. We show that the ensuing entanglement modifies the transition rates and intensities of radiation, as well as introduces a potentially measurable frequency chirp in the SR cascade, the magnitude of which being a function of the separation between the samples. For the dominant SR cascade we find a significant reduction in the duration and an increase of the intensity of the SR pulse relative to the case of a single two-atom SR sample.    

\begin{description}
\item[PACS numbers] 

\end{description}
\end{abstract}

\pacs{33, 98}
                             
\maketitle

\section{\label{sec:Introduction}Introduction}
A group of $N$ two-level atoms in a macroscopically polarized state can be treated as radiating dipoles in the semi-classical approximation. A population-inverted sample can become polarized and emit directionally from a stimulating radiation field or spontaneously from dipole-dipole interactions in the sample geometry. Cooperative emission of a macroscopically polarized sample is known as superradiance (SR) because the emitted power scales as $N^2$. Since its first proposal by Robert Dicke in 1954 \citep{Dicke1954}, there has been continuing extensive research on SR (for reviews see \citep{Gross1982, Siegman1986, Andreev1993, Benedict1996, Allen2012, Scully2009, Manassah2012}). SR has been, and still is, experimentally realized over a wide range of set-ups and conditions. For example, experimental studies of SR were conducted with thermal gases \citep{Gross1976,Yoshikawa2005}, Rydberg atoms at high temperature \citep {Gounand1979, Kaluzny1983}, Bose-Einstein condensates (BECs) \citep{Inouye1999}, quantum dots \citep{Chen2003, Scheibner2007}, semiconductors \citep{Vasil2009}, molecular aggregates in crystals \citep{Malcuit1987}, trapped ions \citep{Devoe1996}, diamond nanocrystals \citep{Bradac2017}, and more. Observational evidence for SR has even been recently found in the interstellar medium (ISM) \citep{RAJABI2016B,Rajabi2016C,Houde2018}.

Different aspects of SR can be theoretically studied using fully quantum mechanical (e.g., master equation \citep{Agarwal1970, Agarwal1971}) or semi-classical (e.g., Maxwell-Bloch equations \citep{Friedberg1972, Macgillivray1976, Haake1979}) treatments of the matter-field system. These studies discuss a wide range of physical systems, from a single SR system composed from only two \citep{Chang1971} or three \citep{Feng2013} atoms to arbitrarily large $N$-atom ($N\gg1$) samples. A common focus of previous theoretical studies was the interactions within a single SR sample, where collective effects such as radiation rate modifications and cooperative frequency shifts \citep{Rohlsberger2010} were comprehensively discussed. 

With the growing interest in understanding the behaviour of coupled quantum systems, e.g., for quantum information purposes and experimental developments for fabricating highly controllable quantum systems such as trapped ions, compound SR systems where two or more SR samples are coupled to one another seem an interesting platform to explore. Central to SR are the so-called Dicke states, which are multi-particle entangled states with important applications for quantum information \citep{Peng2010, Prevedel2009}. Although a SR system can be initially prepared in a Dicke state through coherent pumping, in the absence of a coherent pump the interaction of atoms with their common radiation field can place the system in a Dicke state after a delay that is a function of the characteristic SR time $T_{\mathrm R}$. In this paper, we show how a compound SR system acts as an entangled system where the constituent SR samples cannot be treated independently. This can be important for the experimental generation of larger quantum entangled states.

Another important characteristic of a SR sample is the cooperative shift of the SR resonant frequency, which scales linearly with the number of atoms $N$. The cooperative frequency shift, or $N$-atom Lamb shift, can also be understood within the context of the dipole model. It can be shown that the cooperative radiation decay rate, which is the essence of SR, and the cooperative frequency shift are given, respectively, by the imaginary and real parts of resonant dipole-dipole interaction term within a SR sample \citep{Gross1982,Benedict1996}. This cooperative frequency shift was explored experimentally only recently, and is observed in both near- \citep{Frucci2017} ($R<\lambda$, with $R$ is the sample's dimension and $\lambda$ wavelength of the radiation) and far-field \citep{Meir2014} ($R\gg\lambda$) regimes. The cooperative frequency shift in a single SR sample strongly depends on the spatial distribution of the atoms, and can be concealed in a high density medium \citep{Rohlsberger2010}. 

In this paper, we consider the interaction between macroscopic dipoles associated with individual SR samples separated by large distances $r\gg\lambda$. The inter-sample interaction is therefore dominated by the radiation component of the dipole-dipole interaction term (i.e., not through near-field static-like interactions discussed in the previous paragraph). We show how, at resonance, this interaction not only modifies the radiation decay rates of the compound SR system but also leads to significant shifts in the central resonant frequency. Although these effects arising in the compound SR system are due to a mechanism similar to the ones affecting a single SR sample, i.e., resonant dipole-dipole interactions (albeit involving only the radiation term), their amplitudes are modulated by the separation and orientation of the interacting SR samples composing the compound system.

Some examples of naturally occurring compound SR systems are the maser-harbouring regions of the ISM, where SR can ensue when a threshold for the column density of the inverted population (of molecules, in these cases) is met or exceeded \citep{Gross1976}. The studies of SR in the ISM \citep{RAJABI2016A,RAJABI2016B,Rajabi2016C,RAJABI2016THESIS,Houde2018} suggest that the length-scale of a typical SR sample is relatively small compared to that expected for astronomical masers, with the implication that a region initially hosting a maser must break down into a large number of SR samples when the aforementioned threshold is reached. These SR samples are expected to be approximately simultaneously triggered, and the total intensity of radiation emanating from such a region results from a contribution of the individual SR samples located along the line-of-sight. The previous SR analyses adapted to the ISM  do not discuss possible interactions between neighboring SR samples. In this paper we show how the exchange of photons between different SR samples brings about their entanglement and modifies the overall SR intensity, and leads to a potentially significant frequency shifts due to the large number of molecules composing individual SR samples.

We set up the problem of interacting SR samples forming a compound SR system through a simple (macroscopic) dipole-dipole approximation in Sec. \ref{sec:Analysis}, and provide a solution for the case of two interacting two-atom SR samples (Sec. \ref{sec:two-atom}). In particular, using the dressed state formalism \citep{Cohen1994} we calculate transition rates and intensities (Sec. \ref{sec:intensities}), as well as frequency shifts (Sec. \ref{sec:spectrum}) for the radiation emanating from the resulting compound SR system. We also briefly look at the problem from the standpoint of the uncoupled states basis in Sec. \ref{sec:non-interacting}, which emphasizes the entanglement of the compound SR system. In Sec. \ref{sec:field approach}, we use an alternative approach to estimate the order of magnitude of the frequency shifts for samples composed of $N\gg1$ atoms/molecules, while in Sec. \ref{sec:Discussion} we discuss the implications of our results to SR experiments. We end with a short conclusion in Sec. \ref{sec:Conclusion}

\section{\label{sec:Analysis}Interacting superradiance systems}

Let us consider a SR sample containing $N_1$ two-level atoms arranged such that it can radiate along the $z$-axis through a radiation mode of wave number $k$ in the $+z$ direction. We assume this system (hereafter System 1) to be at some position $z=z_0$ and that an observer is located far in the $+z$ direction where the SR signal is measured.  

We allow for the presence of another SR system (hereafter System 2) composed of $N_2$ atoms also radiating along the $z$-axis but located at $z>z_0$. The Hamiltonian $\hat{H}$ for the combined, two-SR samples system can be written down as

\begin{equation}
    \hat{H}=\hat{H}_1+\hat{H}_2+\hat{V},\label{eq:Hamiltonian}
\end{equation}

\noindent where $\hat{H}_j$, with $j = 1,2$, stand for the (non-interacting) Hamiltonians of  SR Systems 1 and 2, and $\hat{V}$ is the interaction term between the two SR samples. 

When $\hat{V}=0$, the SR systems are left to themselves and will spontaneously emit photons independently at the enhanced rate $T_{\mathrm{R},j}^{-1}$ and intensity $I_{\mathrm{SR},j}$ (proportional to the square of the number of excited atoms in the corresponding sample), as originally described by \citet{Dicke1954}. The eigenstates of a SR system can be defined using $\left|N_{\mathrm{e}},n\right>$, where $0 \leq N_\mathrm{e} \leq N_j$ specifies the number of excited atoms in the SR sample and $n$ the number of photons it emitted in the aforementioned radiation mode. Still neglecting the interaction between the two SR systems, we can generally define an uncoupled basis composed of states $\left|N_\mathrm{e1}, n_1\right>\otimes\left|N_\mathrm{e2}, n_2\right> = \left|N_\mathrm{e1}, n_1; N_\mathrm{e2}, n_2\right>$ for the compound, non-interacting SR system. 

We further simplify our model by making a few assumptions concerning the interaction $\hat{V}$. That is, we consider a photon-mediated dipole-dipole interaction between the two aligned SR samples. Furthermore, this dipole-dipole interaction only involves a far-field radiation component, i.e., we do not consider static-like dipole-dipole interaction (i.e., we assume $kr \gg 1$, with $r$ the relative distance separating the two SR samples). Finally, the aforementioned mode $k$ through which the samples interact is the common SR radiation mode of the samples themselves. We therefore associate to each SR system a macroscopic dipole $\hat{\mathbf{D}}_j$ stemming from the individual microscopic dipoles of the $N_j$ atoms composing the SR samples, such that the interaction contained in Eq. (\ref{eq:Hamiltonian}) is approximated with the simple radiation dipole-dipole term 

\begin{equation}   
\hat{V} = -\alpha\left(r\right) \left[\hat{\mathbf{D}}_1 \cdot \hat{\mathbf{D}}_2 -\left(\mathbf{n}\cdot \hat{\mathbf{D}}_1\right)\left(\mathbf{n}\cdot \hat{\mathbf{D}}_2\right)\right], \label{eq:Interaction}
\end{equation}

\noindent where the coupling strength between the macroscopic dipoles is given by $\alpha(r) = k^2 \cos\left(kr\right) / 4\pi\epsilon_0 r$, with $r=\left|\mathbf{r}\right|$ the relative distance separating them, and $\mathbf{n}=\mathbf{r}/r$ is the unit vector from System 1 to System 2 \citep{Jackson1998}.

With the stated assumptions, taking into account the separation between the two aligned dipoles (with $\mathbf{n} = \mathbf{e}_z$) and the fact that the operators $\hat{D}_{j\pm} =\hat{D}_{jx}\pm i\hat{D}_{jy}$ are associated with the annihilation and creation of photons at the positions of the dipoles, we can then rewrite the interaction term as

\begin{equation}   
\hat{V} = \frac{\alpha\left(r\right)}{2} \left[\hat{D}_{1+} \hat{D}_{2-} e^{-ikr}+\hat{D}_{1-}\hat{D}_{2+} e^{ikr} \right]. \label{eq:Interaction2}
\end{equation}

\noindent Eq. (\ref{eq:Interaction2}) also makes clear that the total number of photons in the compound SR system is conserved through the interaction between the two samples. 

Because of the increased beaming of the radiation emanating from a SR sample compared to that from a simple dipole radiator, Eq. (\ref{eq:Interaction2}) is clearly an idealization that should not be expected to perfectly match the interaction between two SR samples. It will, however, be adequate for our purpose in showing the main consequences resulting from this type of interaction. 

Upon considering the interaction term $\hat{V}$ the new eigenstates of the compound interacting SR system are different from, and consist of combinations of, the aforementioned uncoupled states. For systems with large numbers of atoms, solving the eigenstate and eigenvalue problem for the Hamiltonian of Eq. (\ref{eq:Hamiltonian}) is a complicated task. Here, we further simplify the problem by considering two two-atom SR samples at resonance with each other. 

Although the case of two two-atom SR samples can be solved within the general context of Dicke's SR theory by simply considering four atoms arbitrarily positioned, we choose to analyze it within the framework put forth above in the hope that it will serve as a guide for more complicated systems involving larger numbers of atoms and SR samples, while keeping the analysis as simple as possible. One notable simplification for the two-atom SR samples considered in our analysis is that the macroscopic dipole moments associated with the individual samples have the same value for all transitions. This is not the case when more atoms are involved, as the the dipole moment then varies through the SR cascade \citep{Dicke1954,Dicke1964}. Still, we expect the physical effects discussed here to remain qualitatively unchanged.
 
\subsection{Two two-atom superradiance samples}\label{sec:two-atom}

We first introduce a letter-based notation for the eigenstates of a two-atom SR sample to remove any possible ambiguity with
\begin{eqnarray}
\left|c,0\right> & \equiv &\left|N_{\mathrm{e}} = 2,n = 0\right>\label{state1SR1} \\ 
\left|b,1\right> & \equiv &\left|N_{\mathrm{e}} = 1,n = 1\right>\label{state2SR1} \\  
\left|a,2\right> & \equiv & \left|N_{\mathrm{e}} = 0,n = 2\right>.\label{state3SR1}
\end{eqnarray}

\begin{center}
   \begin{figure}
        \includegraphics[scale=1.]{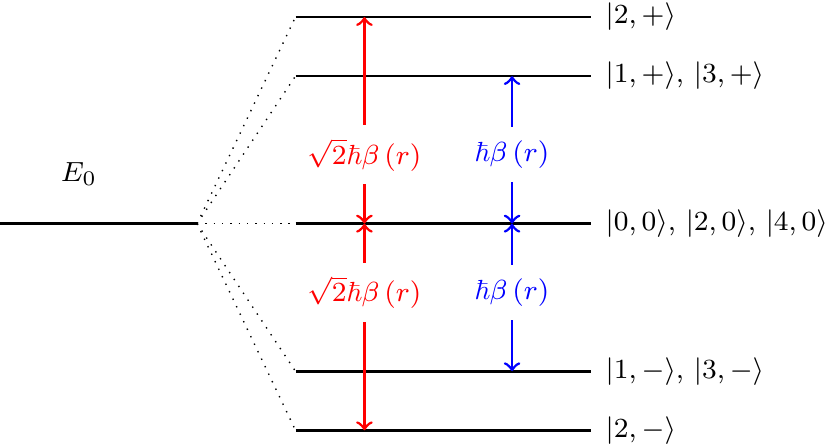}
        \caption{The energy levels associated to the dressed states of the interacting two two-atom SR system. The degeneracy in energy (at $E_0$) of the non-interacting SR systems is partially lifted through the interaction.}
        \label{fig:2SR-energy_levels}
    \end{figure}
 \end{center}

For two two-atom SR samples, assumed to be at resonance with each other, there are nine combined uncoupled states $\left|c,0;c,0\right>$, $\left|c,0;b,1\right>$, $\left|c,0;a,2\right>$, \ldots, $\left|a,2;b,1\right>$, and $\left|a,2;a,2\right>$ that can be used as a starting basis to determine the new eigenstates of the interacting SR system. It is important to note that all of these states are degenerate in energy. However, this degeneracy is partially lifted by the interaction $\hat{V}$, and the new set of dressed eigenstates of the compound system, obtained through a diagonalization of the matrix corresponding to the Hamiltonian of Eq. (\ref{eq:Hamiltonian}), can be shown to be
\begin{eqnarray}
\left|0,0\right> & = & \left|c,0;c,0\right>\label{eq:state0,0}\\
\left|1,\pm\right> & = &\frac{1}{\sqrt{2}}\left(e^{-ikr/2}\left|c,0;b,1\right> \pm e^{ikr/2}\left|b,1;c,0\right> \right)\label{eq:state1pm}\\ 
\left|2,\pm\right> & = &\frac{1}{2}\left(e^{-ikr}\left|c,0;a,2\right> \pm \sqrt{2} \left|b,1;b,1\right> + e^{ikr}\left|a,2; c, 0\right> \right)\label{eq:state2pm}\\ 
\left|2,0\right> & = &\frac{1}{\sqrt{2}}\left(\rule{0in}{2.5ex}e^{-ikr}\left|c,0;a,2\right> - e^{ikr}\left|a,2;c,0\right> \right)\label{eq:state2,0}\\
\left|3,\pm\right> & = &\frac{1}{\sqrt{2}}\left(e^{-ikr/2}\left|b,1;a,2\right> \pm e^{ikr/2}\left|a,2;b,1\right> \right)\label{eq:state3pm}\\
 \left|4,0\right> & = &\left|a,2;a,2\right>,\label{eq:state4,0}
\end{eqnarray}
\noindent where for the states $\left|j,\pm\right>$ and $\left|j,0\right>$, $j$ stands for total number of SR photons emitted by the compound SR system and the $\pm$ and $0$ symbols indicate the state's energy level in relation to that of the non-interacting SR samples. Accordingly, the eigenvalues associated with these dressed states are

\begin{eqnarray}
E\left(j,0\right) & = & E_0,\:\:\:\:\mathrm{for}\:\:j=0,2,4\label{E(j,0)} \\
E\left(1,\pm\right) & = & E_0\pm\hbar\beta\left(r\right)\label{E(1,pm)} \\
E\left(2,\pm\right) & = & E_0\pm\sqrt{2}\hbar\beta\left(r\right)\label{E(2,pm)} \\
E\left(3,\pm\right) & = & E_0\pm\hbar\beta\left(r\right)\label{E(3,pm)}, 
\end{eqnarray}

\noindent where $E_0=4\hbar\omega$ is the degenerate energy level of the non-interacting SR system, with $\hbar\omega$ the energy difference between the two levels of an atom (and therefore the energy difference between the $\left|c\right>$ and $\left|b\right>$, as well as between the    $\left|b\right>$ and $\left|a\right>$ atomic Dicke states). The energy shifts of the dressed states in relation to $E_0$ are determined by 

\begin{equation}
\beta\left(r\right)=\left(2d^2\right)\frac{\alpha\left(r\right)}{2\hbar},\label{eq:Beta}
\end{equation}

\noindent where $d$ is the single-atom dipole moment for the transition under consideration. We intentionally put the term $2d^2$ in parentheses to emphasize the fact that the numerical coefficient (i.e., 2) increases with the number of atoms in the SR samples; this will become important for our discussion in Sec. \ref{sec:Discussion}.  The dressed states energy levels are shown in Fig. \ref{fig:2SR-energy_levels}.  
 
\begin{center}
   \begin{figure}
        \includegraphics[scale=1.]{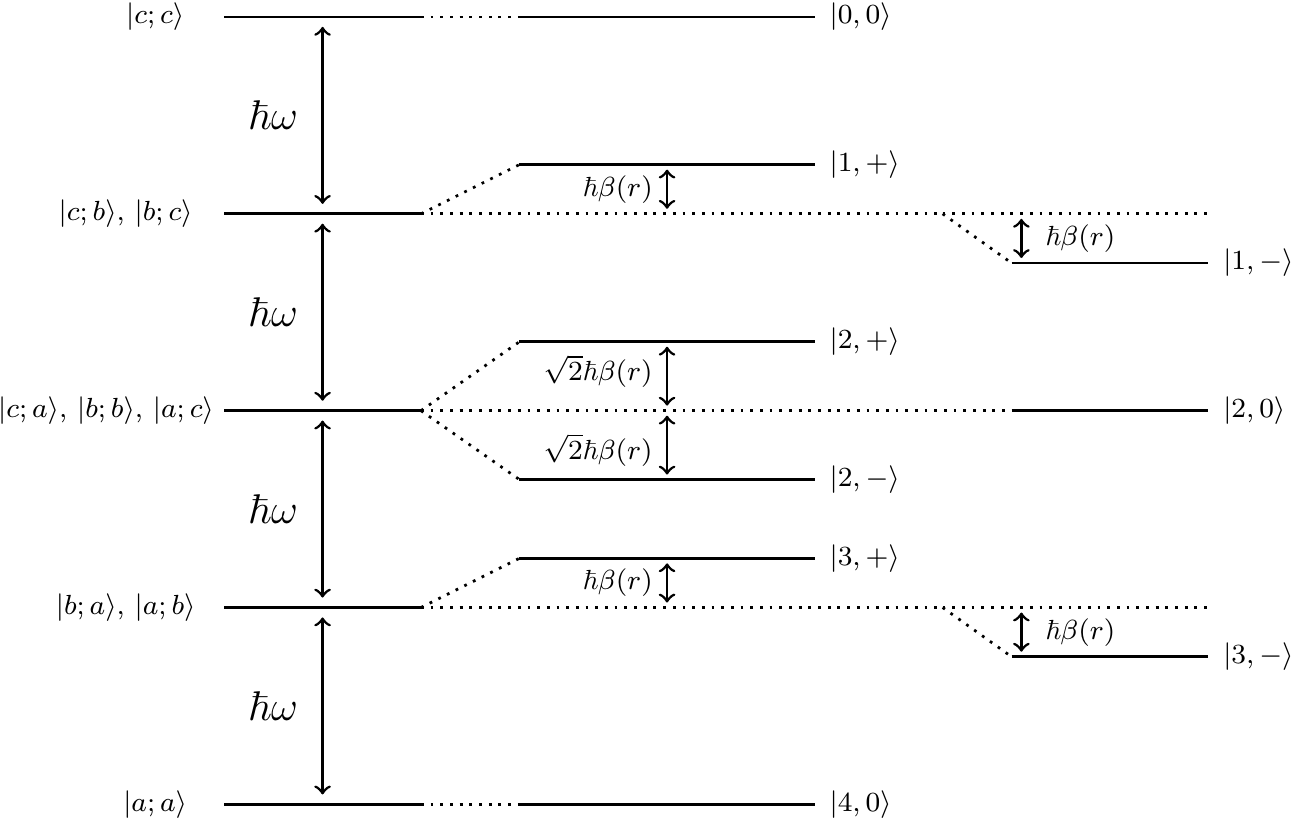}
        \caption{The energy structure of the SR systems relating the dressed states with the atomic Dicke states of the uncoupled system. The Dicke states of the non-interacting SR systems are positioned on the left and the dressed states of the interacting SR system, divided into two groups, on the right of the diagram. In the limit where $kr\rightarrow0$ the states of the first group are symmetric under the permutation of Samples 1 and 2, while those of the second group are anti-symmetric.}
        \label{fig:2SR-cascade}
    \end{figure}
 \end{center}

It is also advantageous to display the energy structure of the SR systems by relating the dressed states with the atomic Dicke states of the uncoupled system (i.e., we omit the number of photons in the uncoupled states to get $\left|c;c\right>$, $\left|c;b\right>$, \ldots, and $\left|a;a\right>$). This is shown in Fig. \ref{fig:2SR-cascade}, where the Dicke states of the non-interacting SR samples are positioned on the left and the related dressed states of the interacting SR system on the right of the diagram. The dressed states were further divided into two groups:
 $\left\{\left|0,0\right>,\left|1,+\right>,\left|2,+\right>,\left|2,-\right>,\left|3,+\right>,\left|4,0\right>\right\}$ and $\left\{\left|1,-\right>,\left|2,0\right>,\left|3,-\right>\right\}$. The dressed states in Eq. (\ref{eq:state0,0})-(\ref{eq:state4,0}) are reminiscent of so-called timed Dicke states \citep{Dicke1964}, while in the limit where $kr\rightarrow0$ they are similar in form to the Dicke states of a single small SR sample. Still when $kr\rightarrow0$, the states of the first group are symmetric under the permutation of Samples 1 and 2 while those of the second group are anti-symmetric, and as the distance between the two SR samples increases the exponential factors in Eq. (\ref{eq:state0,0})-(\ref{eq:state4,0}) mix the symmetries. 
 
\begin{center}
   \begin{figure}
        \includegraphics[scale=1.]{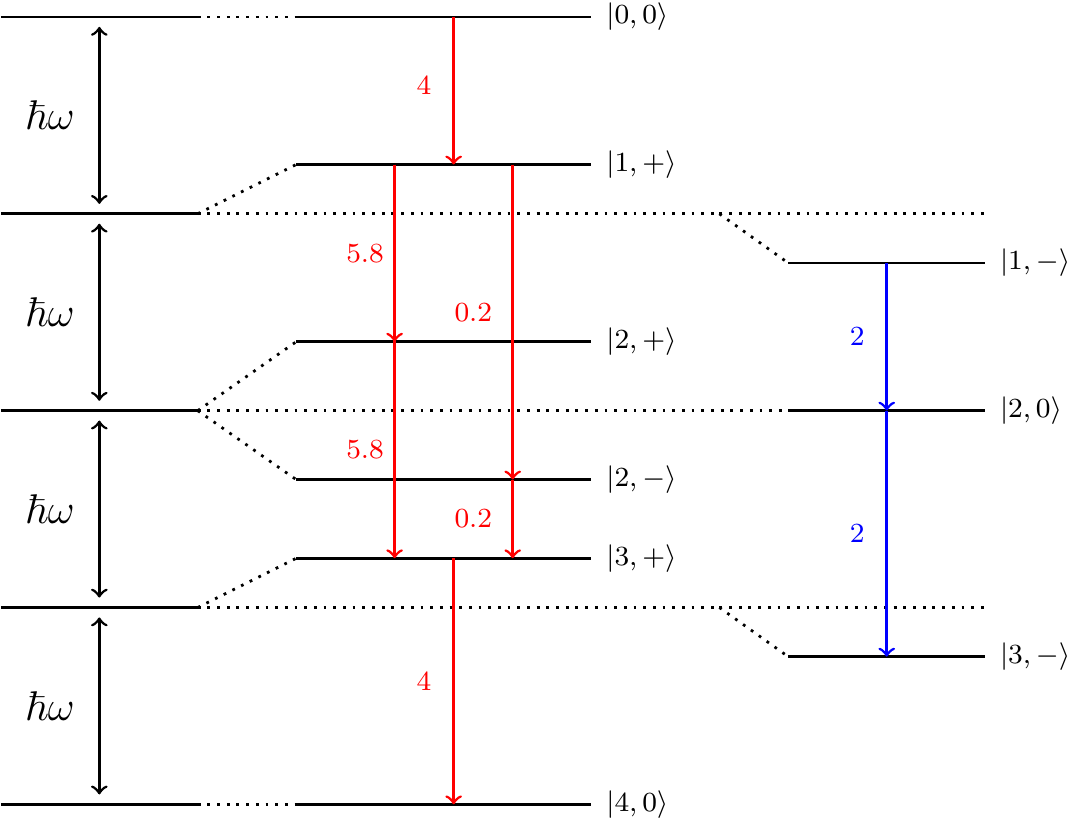}
         \caption{Allowed SR transitions for the different groups of dressed states. The transitions rates, written next to the corresponding vertical arrows, are relative to that of a two-level atom (see Fig. \ref{fig:twoatom-SR.pdf}). No transitions are allowed between the two groups of states.}
        \label{fig:2SR-transitions}
    \end{figure}
 \end{center}

Potential transitions between these states can be identified using the total transverse dipole moment

\begin{equation}
\hat{D}_{\pm}=\hat{D}_{1\pm}e^{\mp ikr/2}+\hat{D}_{2\pm}e^{\pm ikr/2},\label{eq:Dpm}
\end{equation}

\noindent which also exhibits changing symmetries under the permutation of Samples 1 and 2 as a function of $kr$. However, under our set of assumptions we find that transitions will only be allowed between two states contained within the same group and for which the photon number differs by 1. For example, the transition $\left|0,0\right>\leftrightarrow\left|1,+\right>$ is allowed, but $\left|0,0\right>\nleftrightarrow\left|1,-\right>$ and $\left|1,+\right>\nleftrightarrow\left|3,+\right>$ are, as indicated, forbidden. The different allowed SR transitions are shown in Fig. \ref{fig:2SR-transitions}.  



 \subsubsection{Intensities and timescales}\label{sec:intensities}

\begin{center}
   \begin{figure}
        \includegraphics[scale=1.]{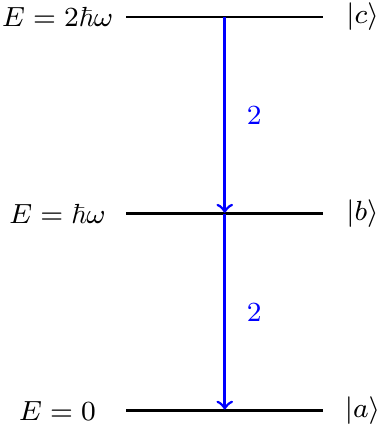}
        \caption{Energy diagram and corresponding transitions for a two-atom SR sample. The transitions rates are relative to that of a two-level atom.}
        \label{fig:twoatom-SR.pdf}
    \end{figure}
 \end{center}

As the study of the possible transitions in Fig. \ref{fig:2SR-transitions} reveals, within the dressed state formalism the two individual SR samples, which by themselves are limited to only two transitions (see Fig. \ref{fig:twoatom-SR.pdf}), are seen to become entangled through their interaction. That is, they transform into a single, larger compound SR system with cascades containing up to four transitions (the two red branches on the left in Fig. \ref{fig:2SR-transitions}). It therefore becomes natural to inquire about the radiation intensity and timescale of these new SR cascades.
 
In Fig. \ref{fig:2SR-transitions}, the rates for all transitions, scaled to that of the two-level atom ($\Gamma = \tau_{\mathrm {sp}}^{-1}$, with $\tau_{\mathrm {sp}}$ the timescale for spontaneous emission), are given next to the corresponding vertical arrows. For example, although there is no rate enhancement for the third group of dressed states (blue arrows on the right in Fig. \ref{fig:2SR-transitions}) compared to a two-atom SR sample (for which all relative transition rates equal 2; see Fig. \ref{fig:twoatom-SR.pdf}), the first and main cascade of states (the leftmost red arrows on the figure) shows an enhancement by a factor of 2 for the first and last transitions of the cascade and $3/2+\sqrt{2} \simeq 2.9$ for the middle two. The second branch of states (middle red arrows; hereafter secondary) shows a markedly reduced transition rate of $3/2-\sqrt{2} \simeq 0.1$ for the middle transitions (still relative to a two-atom SR sample).

We note that the transition rates between the dressed states do not show any dependency on $r$, the distance between the two SR samples. Although this may appear surprising at first, this behaviour can be traced to the fact that we are considering only one radiation mode in our analysis. For example, the inclusion of other radiation modes sharing the same value for $k$ but of different orientations would bring the appearance of phase terms dependent on $kr$.

Focusing on the main and secondary (red) cascades and assuming that the compound SR system is initially in the state $\left|0,0\right>$, the probability $P_j\left(t\right)$ of finding the system in the dressed state $\left|j\right>$ at time $t$ can be calculated using the rates given in Fig. \ref{fig:2SR-transitions}. The intensities of radiation corresponding to the two cascades can then be evaluated with

\begin{equation}
I_\mathrm{SR}= \hbar\omega \sum_j \gamma_{j,j-1} P_j\left(t\right), \label{eq:I_SR}
\end{equation}

\noindent where the summation is on the upper states for all allowed transitions between a state $\left|j\right>$ and the next state $\left|j-1\right>$ down the SR cascade, with $\gamma_{j,j-1}$ the corresponding transition rate \citep{Benedict1996}. In Eq. (\ref{eq:I_SR}) we also assumed that all transitions result in the emission of a photon of energy $\hbar\omega$. Although, as we will soon show, this is not the case for the interacting SR system, this approximation is perfectly adequate for the present discussion.  

\begin{center}
   \begin{figure}
        \includegraphics[scale=0.65]{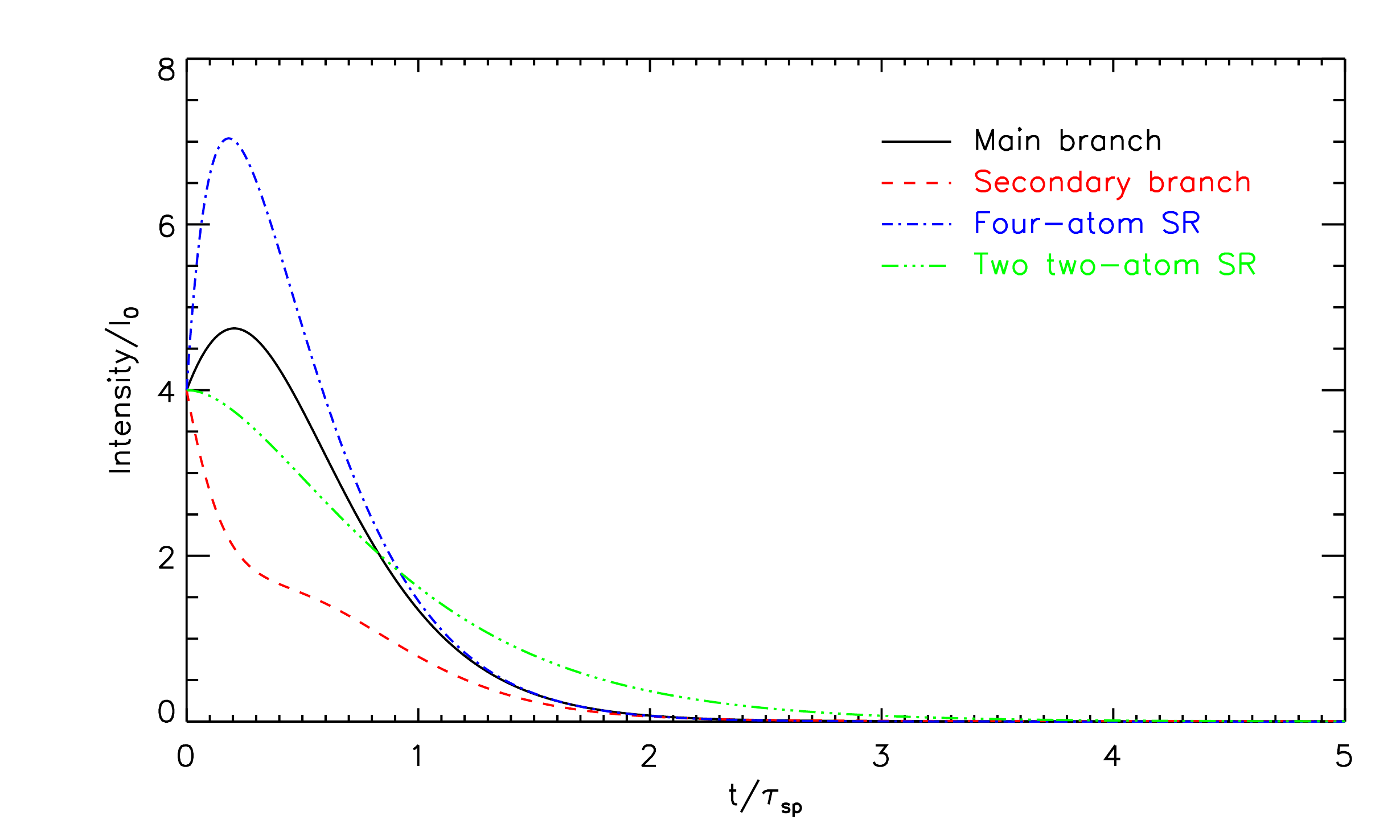}
        \caption{Intensity curves for the main and secondary  branches (solid black and broken red, respectively) of the interacting SR system. For comparison, the compound non-interacting SR system (triple-dotted-broken green), where the intensity is simply twice the intensity of a single two-atom SR sample, and a single four-atom SR sample (dotted-broken blue) are also shown. Time is normalized to $\tau_{\mathrm{sp}}$ and the intensity scaled to $I_0$, the radiation intensity due to spontaneous emission from a single two-level atom.}
        \label{fig:Intensity-plots}
    \end{figure}
    
 \end{center}

In Fig. \ref{fig:Intensity-plots}, the intensities of the main and secondary branches (shown, respectively, in solid black and broken red) are plotted as a function of time using Eq. (\ref{eq:I_SR}). In the figure, time is normalized to $\tau_{\mathrm{sp}}$ and the intensity scaled to $I_0$, the radiation intensity due to spontaneous emission from a single two-level atom. To better gauge the effect of the interaction between the two SR samples, we have also plotted the intensity curves for the compound non-interacting SR system (shown in triple-dotted-broken green), where the intensity is simply twice the intensity of a single two-atom SR sample, and a single four-atom SR sample (shown in dotted-broken blue). The intensities of the single two-atom and four-atom SR samples are calculated in the same manner as for the compound SR system, i.e., by calculating the probabilities of finding the sample in the available states and then using Eq. (\ref{eq:I_SR}).

It is important to note that the interaction between the SR samples not only significantly enhances the transition rates, but also the intensity of the main branch, which is the most probable cascade path for the interacting SR system. The decreased radiation intensity of the secondary branch is also clear as the broken red curve falls below the compound non-interacting SR intensity curve (as well as that for a system of four independent atoms radiating non-coherently; not shown in the figure). On the other hand, the intensity of the main branch of the interacting SR system is noticeably weaker than that of the single four-atom SR sample. Although the entanglement between the two SR samples enhances the transition rates and the intensity of the outgoing radiation, they can never reach those of a sample where all atoms are part of a single four-atom SR sample, as originally described by \citet{Dicke1954}. We expect that these deviations (i.e., the enhancement and reduction relative to, respectively, the two- and four-atom SR samples) to become more significant as the total numbers of atoms and interacting SR systems increase.      

\subsubsection{Frequency shifts}\label{sec:spectrum}

As seen in Figs. \ref{fig:2SR-energy_levels}, \ref{fig:2SR-cascade}, and \ref{fig:2SR-transitions}, the dressed states of the compound interacting SR system span a range of energies, which are shifted by functions of the coupling parameter $\beta\left(r\right)$ relative to the energies of the uncoupled states. Therefore, the cascades of transitions through different branches correspond to sequences of photons of changing frequencies. For example, the main branch corresponds to the following sequence 
\[
\left[\omega - \beta\left(r\right)\right]\rightarrow \left[\omega - \left(\sqrt{2}-1\right)\beta\left(r\right)\right]\rightarrow  \left[\omega + \left(\sqrt{2}-1\right)\beta\left(r\right)\right]\rightarrow  \left[\omega + \beta\left(r\right)\right],
\]
where we note the mirroring of the frequency shifts about $\omega$. We therefore see that the frequency shift through the cascade evolves in a predictable manner with time, as well as with the SR intensity of radiation from the interacting SR system. More precisely, the frequency shift is most negative (at $-\beta\left(r\right)$, when $\beta\left(r\right)>0$) at the start, gradually becomes less negative, crosses zero halfway to become positive, and is maximum (at $+\beta\left(r\right)$) at the end of the cascade. This systematic change in frequency through the cascade is equivalent to a chirp that will frequency modulate the SR signal. 

The importance of this effect evidently depends on the frequency excursion over the SR pulse, in this case $\delta\omega=2\beta\left(r\right)$. However, the shifts in the energy levels of the dressed states relative to those of the unperturbed states and the amplitude of the total frequency excursion are a function of the numbers of atoms and individual SR samples involved in the problem. For example, it can be shown that when $N$ two-atom SR samples are interacting (with the same degree of interaction for all pairs of SR samples) the total frequency excursion is $\delta\omega=2\left(N-1\right)\beta\left(r\right)$. In Section \ref{sec:Discussion}, we discuss an alternative method from Eq. (\ref{eq:Beta}) for estimating $\beta\left(r\right)$.

\subsubsection{The Non-interacting basis}\label{sec:non-interacting}
 
We have so far studied the interaction of SR samples through the dressed states basis. Although the main results leading to the existence of modified transition rates and intensities, and of frequency shifts are more transparent using this formalism, it is also instructive to study the interaction of the two two-atom SR samples with the uncoupled basis composed of the $\left|c,0;c,0\right>$, $\left|c,0;b,1\right>$, $\left|c,0;a,2\right>$, \ldots, $\left|a,2;b,1\right>$, and $\left|a,2;a,2\right>$ states. 

\begin{center}
   \begin{figure}
        \includegraphics[scale=1.]{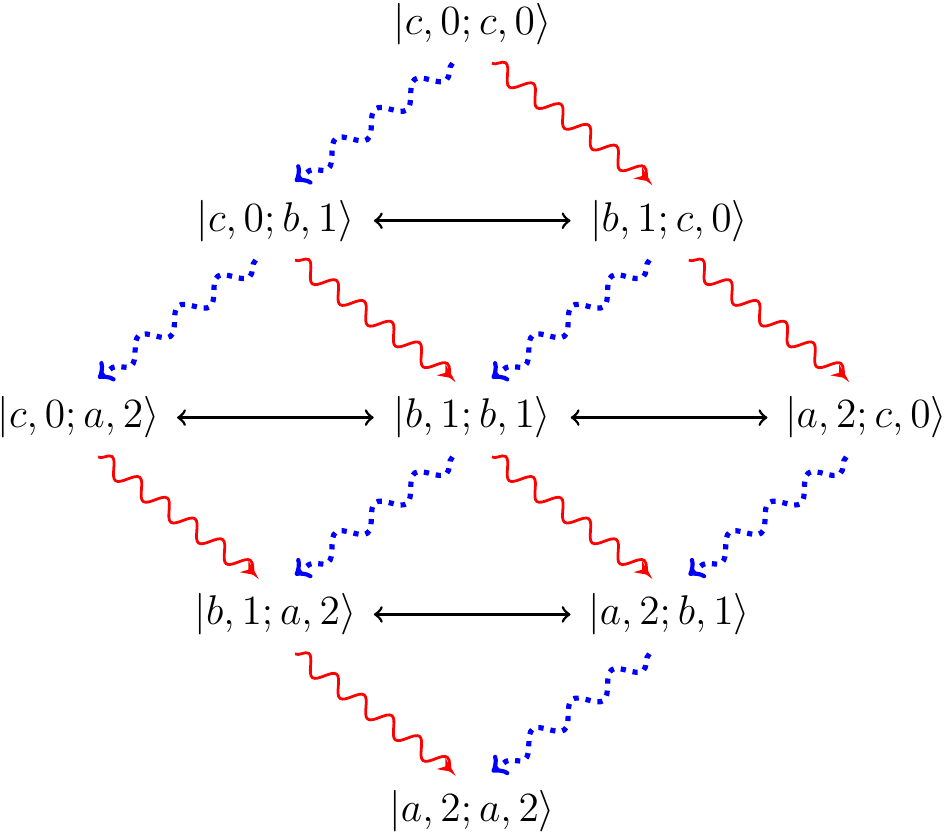}
        \caption{Diagrammatic representation of the SR cascade in the uncoupled basis together with the interaction between the two two-atom SR samples. The dotted blue and solid red curly arrows are for spontaneously emitted SR photons, while the horizontal black double-arrows signify photon exchanges between the two SR samples mediated through their interaction (see Eq. (\ref{eq:Interaction2})).}
        \label{fig:curly-diagram-2twoatom.pdf}
    \end{figure}
 \end{center}  

The SR cascade together with the interaction of Eq. (\ref{eq:Interaction2}) as seen from the uncoupled basis are diagrammatically shown in Fig. \ref{fig:curly-diagram-2twoatom.pdf} (for simplicity, we omit the exponential factors from the states; see Eqs. (\ref{eq:state0,0})-(\ref{eq:state4,0})). In the figure, the dotted blue and solid red curly arrows are for spontaneously emitted SR photons, while the horizontal black double-arrows signify photon exchanges between the two SR samples mediated through their interaction (Eq. (\ref{eq:Interaction2})). We therefore see that as a SR photon is emitted by, say, the second system from the initial $\left|c,0;c,0\right>$ state to the subsequent $\left|c,0;b,1\right>$ state (top dotted blue curly arrow) at the start of the cascade, there is a probability that the cascade will be perturbed by the transfer of a photon between the $\left|c,0;b,1\right>$ and $\left|b,1;c,0\right>$ states.  This photon exchange happens through a Rabi transition of frequency $\left<c,0;b,1\right|\hat{V}\left|b,1;c,0\right>/\hbar$. The SR cascade can then proceed via the different routes through the subsequent levels in the diagram, with or without further Rabi transitions. In the end, four SR photons will have been emitted when the final $\left|a,2;a,2\right>$ state is reached.  It is interesting to note that, within this representation, the two two-atom SR samples do not necessarily emit an equal number of SR photons (i.e., two). Rather, both systems have finite probabilities of emitting anywhere from zero to four SR photons. This underlines the fact that we should resist the temptation of looking at the two two-atom SR samples as independent entities; it is the larger compound and entangled system that emits a total of four SR photons. 

Fig. \ref{fig:curly-diagram-2twoatom.pdf} makes it clear how uncoupled states containing the same number of photons become entangled through the interaction term $\hat{V}$. The modified transition rates and intensities previously discussed are manifestations of these entanglements in the compound system resulting from the exchange of Rabi photons between the two smaller two-atom SR samples.

\section{Coupling of an SR sample to an external radiation field: an alternative approach to the dipole-dipole model}\label{sec:field approach}

As was discussed in Sec. \ref{sec:Analysis}, our choice of a dipole-dipole term for the interaction between two SR samples (see Eqs. (\ref{eq:Interaction}) and (\ref{eq:Interaction2})) is an idealization that cannot perfectly match reality. The dipole-dipole term likely overestimates the solid angle over which the two samples can significantly interact and underestimates the strength of their interaction $\beta\left(r\right)$ when the samples are well aligned with each other. 

An alternative approach can be used to get an approximate estimate for the coupling strength between the two SR samples. That is, instead of considering the presence of two interacting SR systems, we can consider only one such system (System 2) interacting with an external SR radiation field (which, of course, is due to the presence of System 1). In order to preserve the total number of photons (see Eq. (\ref{eq:Interaction2})) the interaction for this model can be written as 

\begin{equation}
\hat{V}=-i\lambda\mathcal{E}_0\left(\hat{D}_{2-}\hat{a}\,e^{ikr}-\hat{D}_{2+}\hat{a}^\dag \,e^{-ikr}\right),\label{eq:interaction-field}
\end{equation}

\noindent where $\lambda<1$ is a coupling coefficient, $\hat{a}$ and $\hat{a}^\dag$ are, respectively, the photon annihilation and creation operators for the SR external radiation field, and (for the one-photon electric field)

\begin{equation}
\mathcal{E}_0=\sqrt{\frac{\hbar\omega}{2\epsilon_0 \mathcal{V}}}, \label{Omega0}
\end{equation}

\noindent with $\epsilon_0$ the permittivity of vacuum and $\mathcal{V}$ the volume of quantization. 

The states of the SR sample are, as before, $\left|c,0\right>$, $\left|b,1\right>$, and $\left|a,2\right>$. For the external SR field the states are $\left|\tau,n\right>\equiv\left|n\right>$, where $n$ is the number of photons in the field and $\tau$ is a label representing the ``atomic" state of the system responsible for the radiation field, which we omit from our notation, for convenience.

Upon applying this model to the two interacting two-atom SR samples (i.e., with System 1 responsible for the external radiation field), the uncoupled basis is given by the nine kets $\left|0;c,0\right>$, $\left|1;c,0\right>$, $\left|2;c,0\right>$, \ldots, $\left|1;a,2\right>$, and $\left|2;a,2\right>$ (the number on the left is for the number of photons  $n$ in the external SR field), and the interaction brings a new set of dressed states. Although not exactly the same, these dressed states have a similar structure to those of Eqs. (\ref{eq:state0,0})-(\ref{eq:state4,0}). Likewise, the interaction between the SR sample and the external radiation field can be visualized in the uncoupled basis with the same diagram as in Fig. \ref{fig:curly-diagram-2twoatom.pdf} (with the appropriate basis, i.e., replacing in the figure $\left|b,1;c,0\right>$ with $\left|1;c,0\right>$, etc.). Accordingly, the eigenvalues of these new dressed states are reminiscent (though not the same) as those found in Eqs. (\ref{E(j,0)})-(\ref{E(3,pm)}) and suggest that we can make the following approximate substitution

\begin{equation}
\hbar\beta\left(r\right)\rightarrow\lambda\sqrt{2}d\mathcal{E}_0. \label{eq:beta-E0}
\end{equation}

It is important to note that in this relation $\sqrt{2}d$ stands for the macroscopic dipole moment of the two-atom SR sample (see Fig. \ref{fig:twoatom-SR.pdf}). That is, if the SR sample consisted of $M$ two-level atoms, then $\sqrt{2}d$ would be replaced by $\sqrt{M}d$ (more precisely, this would be the case for the first transition at the start of the cascade \citep{Dicke1964}). 

Eq. (\ref{eq:beta-E0}) allows us to circumvent some shortcomings of the dipole-dipole interaction model we used in the previous sections to get an approximate value for the magnitude of the maximum frequency shift in the SR cascade of the main branch. We could then surmise that for $N$ interacting $M$-atom SR samples

\begin{equation}
\hbar\delta\omega\sim2\lambda\left(N-1\right)\sqrt{M}d\mathcal{E}_0. \label{eq:shift}
\end{equation}

The relative importance of $\delta\omega$ is thus related to the one-photon Rabi frequency  of the system ($\sim d\mathcal{E}_0/\hbar$), and will depend on the number of interacting samples and the number of atoms they contain.

\section{Discussion}\label{sec:Discussion}

In Section \ref{sec:two-atom}, we discussed the frequency shifts and intensity modifications arising from the interaction between two SR samples, each composed of two emitters. The magnitude of the frequency shifts is modulated by the distance between the interacting SR samples and their relative alignment. Although we simplified our discussion using a dipole-dipole model, where the macroscopic dipoles associated to each SR sample are perfectly aligned, here we propose a physical system and experiment through which our results could be tested. 

\begin{center}
   \begin{figure}
        \includegraphics[scale=0.5]{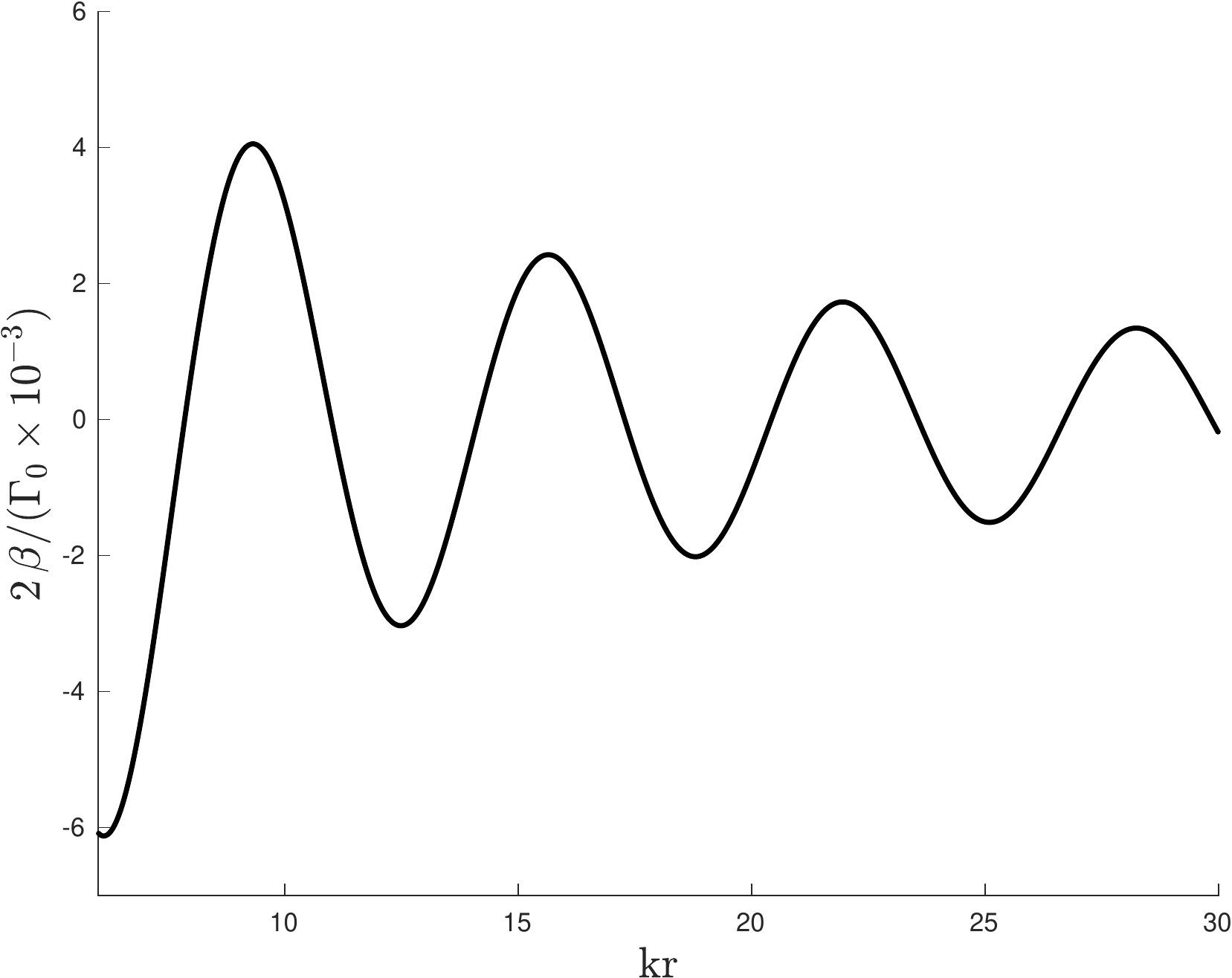}
        \caption{Predicted excursion about the central frequency resulting from the interaction between two two-${\mathrm{Ba}_{138}}^+$ SR samples. The frequency shift is scaled to $ \Gamma_0 \times 10^{-3}$, where $\Gamma_0 = 1/\tau_{\mathrm{sp}}\sim \, 100$ MHz is the natural line width of the 493 nm transition of ${\mathrm{Ba}_{138}}^+$ used in this example. The interaction between the two SR samples could produce a shift as large as 150 kHz at $kr \sim\, 25$.}
        \label{fig:freshift_ba}
    \end{figure}
 \end{center}

As mentioned earlier in Section \ref{sec:Introduction}, SR has been observed for two trapped ions where the collective decay rate is studied as a function of the inter-ion distance \citep{Devoe1996}. Such a system, which allows for the precise spacing and preparation of the emitters, and in general SR samples, seems a perfect candidate for the study of interacting SR samples. In \citet{Devoe1996}, two ${\mathrm{Ba}_{138}}^+$ ions were prepared in a Dicke state using a coherent laser excitation. Their experimental set-up allowed for the perfect control of the phase of the dipoles induced in each ion, with the resulting occurrence of SR for in-phase preparations. Now assume two such SR samples could be prepared, where two ions forming a SR sample are spatially separated by $R$, e.g., $R \sim 0.5 \, \mu$m or $kR = 2\pi \left(R / \lambda \right) \sim \, 6$ for a typical optical transition, and the two two-ion SR samples were placed apart by a distance $r > R$, e.g., $2 \, \mu$m or $kr \sim\, 25$, along a chain. Different properties of the resulting compound SR system (consisting of the two two-ion samples) such as the collective decay rates and frequency shifts could be studied as a function of $r$ and the relative alignment of the two-ion samples. In Figure \ref{fig:freshift_ba} we show the predicted excursion about the central frequency of the compound SR system composed of two aligned two-${\mathrm{Ba}_{138}}^+$ SR samples plotted as a function of $kr$, using Eq. (\ref{eq:Beta}). The angular frequency shift axis is scaled to $ \Gamma_0 \times 10^{-3}$, where $\Gamma_0 = 1/\tau_{\mathrm{sp}}\sim \, 100$ MHz is the natural line width of the 493 nm transition of ${\mathrm{Ba}_{138}}^+$ coupling the two SR samples in this example \citep{Devoe1996}. As seen in the figure, the interaction between the two SR samples could produce a shift as large as 150 kHz for $kr \sim\, 25$. This shift, although very small relative to the natural line width, can be detected by measuring the shift of the line center \citep{Meir2014}. 

In addition one could also study the radiation rate enhancements for the system of two two-${\mathrm{Ba}_{138}}^+$ samples. To do so, one could first measure the collective radiation rate for the samples composed of only two ${\mathrm{Ba}_{138}}^+$ ions as a function of inter-ion distance, similarly to what was done in the \citet{Devoe1996} experiment. The inter-ion distance $R_{\mathrm{max}}$ for which the emission rate is maximum could thus be determined. Then, the experiment could be repeated for a system of two two-${\mathrm{Ba}_{138}}^+$ samples, where for each sample the inter-ion distance is set to $R_{\mathrm{max}}$ while the separation $r$ between the samples was varied. Our model suggests that the collective radiation rates of the compound SR system could be significantly different from the single two-ion or single four-ion samples. More precisely, depending on the collective state of the compound system one can expect superradiance or a decrease in the intensity of radiation. The maximum emission rate of the compound system scaled to that of the two-ion sample could be as large as $\simeq2.9$ (from Fig. \ref{fig:2SR-transitions} and the corresponding discussion in Sec. \ref{sec:intensities}).

In Section \ref{sec:Introduction}, we mentioned that in the ISM the small size of a single SR sample relative to that of the region where the SR requirements can be met implies the simultaneous occurrence of a large number of SR systems \citep{RAJABI2016B, Rajabi2016C}. For example, in the case of the 6.7 GHz methanol and 22 GHz water spectral lines in star-forming regions, it was found that (non-interacting) SR samples could be at least $\sim10^5$ cm in length in these sources (a SR characteristic timescale $T_{\mathrm R}$ on the order of one to many hours) \citep{Rajabi2016C}. Considering the fact that a typical astronomical maser region, where these SR signals were identified, could have lengths on the order of 100 astronomical units (AU; $1\;\mathrm{AU}\simeq1.5\times10^{13}$ cm) \citep{Gray2012}, it follows that a staggeringly large number of SR samples could be contained within a single region.  It is therefore reasonable to expect widespread interaction between separate samples. Such interactions will result in significant frequency shifts in SR regions where bursts of radiation will most likely exhibit a drift in their central SR frequency. The resulting frequency chirps should be readily measurable.

\section{Conclusion}\label{sec:Conclusion}

We considered the interaction between distinct superradiance (SR) systems and use the dressed state formalism to solve the case of two interacting two-atom SR samples at resonance. We showed that the ensuing entanglement modifies the transition rates and intensities of radiation, as well as introduces a potentially measurable frequency chirp in the SR cascade, the magnitude of which being a function of the separation between the samples. For the dominant SR cascade we find a significant reduction in the duration and an increase of the intensity of the SR pulse relative to the case of a single two-atom SR sample.    

\section{Acknowledgements}\label{sec:Acknowledgements}

We are grateful to Dr. Ana Asenjo Garcia for reading the initial draft of this manuscript and providing helpful comments. M. H.'s research is supported by the Natural Science and Engineering Research Council of Canada Discovery Grant RGPIN-2016-04460 and the Western Strategic Support for Research Accelerator Success.

\bibliography{broadband}

\end{document}